\documentclass[a4paper,conference]{IEEEtran}
\IEEEoverridecommandlockouts
\usepackage{cite}
\usepackage{amsmath,amssymb,amsfonts}
\usepackage{algorithmic}
\usepackage{graphicx}
\usepackage{textcomp}
\usepackage{xcolor}
\usepackage[inline]{enumitem}
\usepackage{subcaption}
\usepackage[square,numbers,sort&compress]{natbib}
\usepackage{tikz}
\usetikzlibrary{shapes.geometric, calc, intersections, patterns}
\usepackage{pgfplots}
\pgfplotsset{compat=newest}
\pgfplotsset{plot coordinates/math parser=false} 
\usepgfplotslibrary{polar, fillbetween}

\DeclareMathOperator{\floor}{floor}	

\newcommand{\SNR}{\mbox{SNR}}
\newcommand{\PRF}{\mbox{PRF}}

\def\arXivPrint{1}  

\begin{document}

\title{Analysing Multibeam, Cooperative, Ground Based Radar in a Bistatic Configuration%
}

\author{\IEEEauthorblockN{Pepijn B. Cox and Wim L. van Rossum}
\IEEEauthorblockA{Radar Technology, TNO\\
The Hague, The Netherlands \\
Email: \{pepijn.cox, wim.vanrossum\}@tno.nl}}

%
%
\ifx\arXivPrint\undefined\else

\makeatletter
\twocolumn[{
\vspace{2cm}
This paper has been accepted for publication at the

\vspace{1cm}
\centerline{\textbf{\huge{ 2020 IEEE International RADAR Conference}}}

\vspace{5cm}
978-1-7281-6813-5/20/\$31.00 \textcopyright 2020 IEEE \\
\noindent\hspace*{0.5cm} DIO 10.1109/RADAR42522.2020.9114620

\vspace{1cm}
\textbf{Citation}\\
P.B. Cox and W.L. van Rossum, ``\@title,'' in \textit{Proceedings of the 2020 IEEE International RADAR Conference}, pp 912-917, Washington DC, USA, May 2020.

\vspace{1cm}
\textbf{IEEE Xplore URL}\\
\url{https://ieeexplore.ieee.org/document/9114620}

\definecolor{commentcolor}{gray}{0.9}
\newcommand{\commentbox}[1] {\colorbox{commentcolor}{\parbox{\linewidth}{#1}}}

\vspace{4cm}
\commentbox{
	\vspace*{0.2cm}
	\hspace*{0.2cm}More papers from P.B. Cox can be found at\\~\\
	\centerline{\large{\url{https://orcid.org/0000-0002-8220-7050}}}
	\vspace*{0.0cm}
}

\vspace{3cm}
\textcopyright 2020 IEEE. Personal use of this material is permitted. Permission from IEEE must be obtained for all other uses, in any current or future media, including reprinting/republishing this material for advertising or promotional purposes, creating new collective works, for resale or redistribution to servers or lists, or reuse of any copyrighted component of this work in other works.
}]
\clearpage
\makeatother

\fi
%
%

\maketitle

\begin{abstract}
Recent advances in digital beam forming for phased arrays in combination with digital signal processing should enable the development of multibeam radar in a bistatic configuration. In the bistatic setting, the pulse travelling outward from the transmitter should be followed or ``chased'' by the receiver. During transmission, depending on the location of the transmitter, receiver, and pulse, the number of digital beams and their location at the transmitter vary. In this paper, we analyse the geometrically depending number of digital beams and the beam switching rate of the receiver needed for pulse chasing. In addition, we derive the \emph{pulse repetition frequency} ($\PRF$) for the bistatic configuration based on the desired detection range. It is shown that the $\PRF$ in the bistatic case can be increased compared to its monostatic counterpart when the distance between the transmitter and the receiver is increased. Our results are applied on the scenario of an air traffic control radar to show the feasibility of a multibeam, ground based bistatic surveillance radar. It will be demonstrated that the maximum $\PRF$ can almost be doubled and an adaptive sensing and tracking paradigm can lead to a maximum of 64 simultaneous receiver beams for the bistatic surveillance and tracking setting.
\end{abstract}


\IEEEpeerreviewmaketitle

\vspace*{-4mm}
\section{Introduction}

The setting of non co-located radar transmitter(s) and receiver(s), i.e., the \emph{bistatic} or \emph{multistatic} radar configuration~\cite{Hanle1986,Jackson1986,Willis2004,Wills2008}, dates back to the second world war with the Klein Heidelberg radar~\cite{Griffiths2010}. The non co-located radar can either use transmitters of opportunity -- e.g., FM radio, analogue TV, DVB-T~\cite{Griffiths2010,Howland2005,Samczynski2012} -- which  is known as \emph{passive} multistatic radar or the radar can use dedicated transmitters also known as \emph{active} multistatic radar. In the active setting, the transmitter(s) and receiver(s) can operate in a \emph{cooperative} fashion by exchanging information, such as trajectory of the pulse, waveform type, frequency, etc., to increase the overall accuracy.

In this paper, we analyse the active, cooperative bistatic configuration for ground based radar. More specifically, in this setting, we analyse \emph{pulse chasing}, i.e., a rapid and efficient search of a desired volume of space whereby the receiver antenna will follow or ``chase'' the transmitted pulse during its travel outward from the transmitter antenna~\cite{Purdy2001}. Pulse chasing in a bi- or multistatic configuration has several advantages compared to monostatic radar~\cite{Matsuda2005,Samczynski2012}: 
\begin{enumerate*}[label=\roman*)]
  \item increased detection of targets with small monostatic \emph{radar cross section} (RCS), including stealth targets;
	\item increased detection range into a certain direction, as the detection coverage area changes shape (see Fig.~\ref{fig:ranges of bistatic conf}); and
	\item increased accuracy in position and velocity of the target, when combined with the monostatic channel.
\end{enumerate*}
On the other hand, utilizing efficient pulse chasing requires:
\begin{enumerate*}[label=\roman*)]
	\item rapid receiver beam steering or switching when the pulse is travelling outward and
	\item accurate information of the transmitted beam, i.e., space-time synchronization, via a communication channel, e.g., via a dedicated coaxial cable.
\end{enumerate*}
These demands seem feasible with the recent advances in digital beam switching and digital signal processing.

For pulse chasing, we identify the following four operating mechanisms:
\begin{enumerate*}[label=(\arabic*)]
  \item (analog) pulse chasing using a single, wide beam with time-varying scanning angle~\cite{Hanle1986,Jackson1986},
	\item multiple, simultaneous digital beamformers at the receiver, i.e., \emph{multibeam}, covering the entire to-be-travelled area of the pulse at once,
	\item multibeam with time-varying scanning angle~\cite{Matsuda2005} to cover the pulse only,
	\item multibeam with digitally switching (avoiding translation of the individual beams) covering only the pulse.
\end{enumerate*}
Pulse chasing with one single wide beam will significantly decrease the \emph{signal-to-noise ratio} (SNR)~\cite{Matsuda2005}. For multibeam pulse chasing using a time-varying scanning angle, a target response will pass through multiple beams introducing unwanted fluctuations in target amplitude and phase. These unwanted effects are not present in case 2 and 4 and, therefore, in this paper, we will focus on multibeam with beam switching, i.e., operating mechanisms 2 and 4. We analyse the trade-off between the decreased number of beams of the switched multibeam strategy compared to covering the entire to-be-travelled area on the one hand and the accompanied additional hardware requirements of beam switching and time synchronization on the other hand. 

Hence, in this paper, we analyse multibeam pulse chasing for cooperative, ground based bistatic radar using digital phased arrays. The contributions of this paper are:
\begin{itemize}
	\item First \emph{3 dimensional} (3D) analysis for bistatic pulse chasing (to the authors knowledge).
	\item Calculation of the number of simultaneous digital beams for the receiver and the associated beam switching rates.
	\item Definition of the bistatic \emph{pulse repetition frequency} (PRF).
\end{itemize}

The paper is organized as follows. In Section~\ref{sec:Bistatic Geometry}, the 3D geometry of the bistatic configuration is defined and analysed. Pulse chasing and the digital beam former are explained in Section~\ref{sec:pulse chasing}. In Section~\ref{sec:pulse chasing results}, we analyse these notions using a scenario based on an air traffic control radar in the bistatic configuration followed by the conclusions in Section~\ref{sec:conclusion}.


\section{Bistatic Geometry} \label{sec:Bistatic Geometry}

In this section, the 3D geometry of pulse chasing for ground based, bistatic radar is described. In Section~\ref{subsec:3d coordinate system}, the 3D coordinate system is introduced, followed by the bistatic range equation, in Section~\ref{subsec:bistatic range}. The bistatic blind range and the bistatic pulse repetition frequency are discussed in Section~\ref{subsec:blind range} and~\ref{subsec:PRF range}, respectively. These definitions are used for the pulse chasing setting in Section~\ref{sec:pulse chasing}.

The following discussion is far from complete on all aspects of the geometry for bistatic radar. The interested reader is referred to, e.g.,~\cite{Hanle1986,Jackson1986,Willis2004,Wills2008}.

\subsection{3D Coordinate System} \label{subsec:3d coordinate system}

For the ground based radar, the north-east-down coordinate frame is applied. We assume that the earth is flat and we do not take any height differences into account, e.g., trees or hills. The transmitter and receiver beams are assumed narrow with a rectangular shape without side lobes. The origin frame is denoted as $\mathcal{O}$ and the transmitter and receiver frames are defined by the $(x,y,z)$-coordinates $\mathcal{T}_x\in\mathbb{R}^3$ and $\mathcal{R}_x\in\mathbb{R}^3$, respectively. It is assumed that the transmitter $\mathcal{T}_x$ and receiver $\mathcal{R}_x$ frames are not rotated with respect to each other nor with respect to the origin $\mathcal{O}$. For the sake of simplicity, we assume that the distance between the transmitter and the receiver is $L\in\mathbb{R}^+_0$ and the transmitter is located at $\mathcal{T}_x=[-L/2~0~0]^\top$ and the receiver is located at $\mathcal{R}_x=[L/2~0~0]^\top$.

The radar beam position is given in a spherical coordinate system with an azimuth $\varphi\in\mathbb{R}$, an elevation $\beta\in\mathbb{R}$, and the distance of the travelled radar beam $r\in\mathbb{R}_0^+$. The spherical coordinate system can be converted to the Cartesian coordinates in a north-east-down coordinate frame by the function $\mathcal{P}:\mathbb{R}^3\rightarrow\mathbb{R}^3$
\begin{equation}
\mathcal{P}(r,\varphi,\beta)=\left[\begin{array}{ccc} \!\!
r\cos\varphi\cos\beta & \! r\sin\varphi\cos\beta & \!-r\sin\beta \!\!
\end{array}\right]^\top\!\!\!\!.
\label{eq:polar to cartesian}
\end{equation}
Following, the location of a transmitted pulse as a function of time can be derived using the defined coordinate frames. For simplicity, the radar pulse is represented by a box (polyhedron) given by the 3dB beamwidth in azimuth $BW_{T,a}\in\mathbb{R}^+$ and elevation $BW_{T,e}\in\mathbb{R}^+$. Assuming that the transmission of the pulse starts at $t=0$ with pulse length $\tau_p\in\mathbb{R}^+$ for the transmitted azimuth $\varphi_T$ and the transmitted elevation $\beta_T$, the coordinates of the vertices of the pulse box in the origin frame $\mathcal{O}$ for $t\in[\tau_p~\infty)$ are
\begin{equation}
\begin{aligned}
\mathcal{S}_i &\!=\! \mathcal{T}_x \!+\! P\!\left( r_{p_e},~\varphi_T\!+\!(-1)^{g_i}\varphi_{BW},~\beta_T\!+\!(-1)^{h_i}\!\beta_{BW}  \!\right)\!, \\
\mathcal{S}_j &\!=\! \mathcal{T}_x \!+\! P\!\left( r_{p_s},~\varphi_T\!+\!(-1)^{g_j}\varphi_{BW},~\beta_T\!+\!(-1)^{h_j}\!\beta_{BW}  \!\right)\!, 
\end{aligned}
\label{eq:3Dbox}
\end{equation}
for the pulse trailing surface defined by vertices $i=\{1,\ldots,4\}$ and the pulse leading surface defined by vertices $j=\{5,\ldots,8\}$ with $r_{p_e}=c(t-\tau_p-\delta t)$, $r_{p_s}=c(t+\delta t)$, $\varphi_{BW}=\frac{1}{2}BW_{T,a}+\delta\psi_a$, $\beta_{BW}=\frac{1}{2}BW_{T,e}+\delta\psi_e$, $g_i=\floor(i/2-1)$, and $h_i=\floor((i+1)/2)$ where $\delta t\in\mathbb{R}^+_0$ models potential timing inaccuracies, $\delta\psi_a,\delta\psi_e\in\mathbb{R}^+_0$ models (measurement) inaccuracies in the azimuth or elevation direction, respectively, and $c$ is the speed of light in vacuum. In Section~\ref{sec:pulse chasing}, pulse chasing is discussed based on the transmitted pulse represented by the box~\eqref{eq:3Dbox}.

\subsection{Bistatic Range} \label{subsec:bistatic range}

The radar range equation is used to determine the largest volume that can be searched by a radar~\cite{Willis2004}. The bistatic range equation is equivalent to the monostatic counter part. However, in the bistatic case, the transmitter-to-target range $R_T\in\mathbb{R}^+$ and the receiver-to-target range $R_R\in\mathbb{R}^+$ are not equivalent, because the transmitter and receiver are not co-located. The signal-to-noise power ratio contours, based on the radar range equation, reads as~\cite{Willis2004}
\begin{equation}
\SNR = \frac{k}{R_T^2R_R^2},
\label{eq:SNR range}
\end{equation}
with bistatic radar constant $k$
\begin{equation*}
k = \frac{P_TG_TG_R\lambda^2\sigma_BF^2_TF^2_R}{(4\pi)^3KT_sB_nL_TL_R},
\end{equation*}
where $P_T\in\mathbb{R}^+$ is the transmitter power, $G_T,G_R\in\mathbb{R}^+$ are the transmit or the receive antenna power gain, $\lambda\in\mathbb{R}^+$ is the wavelength, $\sigma_{\mathrm{bi}}\in\mathbb{R}^+$ is the bistatic target \emph{radar cross section} (RCS), $F_T,F_R\in\mathbb{R}^+$ are the propagation factor for the transmitter-to-target path or for the target-to-receiver path, $K$ is the Boltzmann's constant, $T_s\in\mathbb{R}^+$ is the receive system noise temperature, $B_n\in\mathbb{R}^+$ is the noise bandwidth of receiver after processing, and $L_T,L_R>1$ are the transmit and receive system losses.

Note that, similar to~\cite{Willis2004,Wills2008,Matsuda2005,Samczynski2012}, we assume an invariant bistatic radar constant $k$ in~\eqref{eq:SNR range} by neglecting that:
\begin{enumerate*}[label=\roman*)]
	\item the propagation factors $F_T,F_R$ are dependent on the distances $R_T$ and $R_R$,
	\item the transmit or receive antenna power gain $G_T,G_R$ are dependent on the azimuth and elevation angle,
	\item the post-processed bandwidth noise $B_n$ can be dependent on many time and spatial factors, and
	\item the bistatic RCS $\sigma_{\mathrm{bi}}$ can vary for different aspect angles of the target.
\end{enumerate*}

For a given bistatic radar constant $k$ and a minimum $\SNR$, the contour lines for the maximum bistatic range $R_{bi}=\sqrt{R_TR_R}$ can be determined. The contours are one form of the ovals of Cassini~\cite{Jackson1986,Willis2004,Wills2008}. The coordinates of the Cassini oval surface with focal points $\mathcal{T}_x=[-L/2~0~0]^\top$ and $\mathcal{R}_x=[L/2~0~0]^\top$ can be written as a function in the spherical coordinates $(\vartheta,\varpi)\in[0~2\pi)\times[0,\pi)$ in the origin frame $\mathcal{O}$ for $L\leq2\sqrt{R_TR_R}$ given by
\begin{equation}
\mathcal{R}_{cas}(\vartheta,\varpi) = \frac{L}{2}\sqrt{C(\vartheta)}\left[ \begin{array}{c}
	\cos(\vartheta)\\
	\cos(\varpi)\sin(\vartheta)\\
	-\sin(\varpi)\sin(\vartheta)
\end{array} \right],
\label{eq:Cassini contour}
\end{equation}
where
\begin{equation*}
C(\vartheta)=\cos(2\vartheta)+\sqrt{\frac{16R_T^2R_R^2}{L^4}-\sin^2(2\vartheta)}.
\end{equation*}
The function $C(\vartheta)$ reaches its maximum at $\vartheta=k\pi$ with $k\in\mathbb{Z}$ and its minimum at $\vartheta=(k+0.5)\pi$ with $k\in\mathbb{Z}$~\cite{Yates1952}. In case that $L>2\sqrt{R_TR_R}$, the Cassinian surface splits into two separate shapes, which is not in our area of interest. Note that the Cassinian surface~\eqref{eq:Cassini contour} is symmetric over the x-axis. An example of the bistatic detection contour is given in Fig.~\ref{fig:ranges of bistatic conf}.

\subsection{Bistatic Eclipsing} \label{subsec:blind range}

For pulsed radar in a bistatic configuration, eclipsing occurs when the direct signal from the transmitter to the receiver overlaps with the reflected signal from the target. Eclipsing exists if the target is within the following ellipse%
\footnote{An ellipse satisfying $R_R+R_T=2a$ with the foci on coordinates $[\pm l,~0,~0]^\top$ and semi-major axis $a$ will have a semi-minor axis $b=\sqrt{a^2-l^2}$ and the elliptical surface coordinates are described by the function $\mathcal{R}_{ell}(\vartheta,\varpi)=\left[ a\cos(\vartheta)\cos(\varphi),~b\cos(\vartheta)\sin(\varphi),~b\sin(\vartheta)\right]^\top$ for the spherical coordinates $(\vartheta,\varpi)\in[0,2\pi)\times[0,2\pi)$ from the origin $\mathcal{O}$.}
\begin{equation}
R_T + R_R \leq L+c\tau_p.
\label{eq:eclipsing}
\end{equation}
In the analysis in Section~\ref{sec:pulse chasing results}, we assume that the bistatic radar configuration is blind for any target within the eclipsing range~\eqref{eq:eclipsing}. This implies that we do not take forward scattering into account. Bistatic eclipsing is displayed in Fig.~\ref{fig:ranges of bistatic conf} as the blank oval near the x-axis. 

\subsection{Bistatic $\PRF$} \label{subsec:PRF range}

Choosing the maximum pulse repetition frequency in the bistatic configuration is different than in the monostatic case where $\PRF_{\mathrm{mo}}\leq\frac{c}{2R_{\mathrm{max}} + c\tau_{p}}$. Focusing on the unambiguous solution, the leading and trailing edge of the transmitted pulse from the transmitter-to-target-to-receiver will follow an elliptical shape
\begin{subequations} \label{eq:PRF ellipse}
\begin{align}
&\mbox{leading:}\hspace*{0.5cm}  &R_T+R_R&= L+c\left(\frac{1}{\PRF_{\mathrm{bi}}} \right), \label{eq:PRF ellipse max} \\
&\mbox{trailing:} &R_T+R_R&= L+c\left(\frac{1}{\PRF_{\mathrm{bi}}}-\tau_p \right), \label{eq:PRF ellipse min}
\end{align}
\end{subequations}
similar to the eclipsing range in Section~\ref{subsec:blind range}. To determine the maximum bistatic $\PRF_{\mathrm{bi}}$, the leading wave of the next pulse should not reach the receiver before the trailing edge of the reflected signal of the previous pulse reaches the receiver, i.e.,
\begin{equation}
\PRF_{\mathrm{bi}}\leq\frac{c}{2 \Delta R_{\mathrm{bi}}\!+\! c\tau_p } = \frac{c}{\sqrt{\!L^2\!+\!4R_TR_R}\!-\!L\!+\!c\tau_p},
\label{eq:bistatic PRF}
\end{equation}
for $L\leq2\sqrt{R_TR_R}$, where $\Delta R_{\mathrm{bi}}=L/2 ( \sqrt{\!C(k\pi)}\!-\!1)$ is the distance between the receiver and the maximum detection range in the positive $x$ direction for $k\in\mathbb{Z}$. In the appendix, it is shown that condition~\eqref{eq:bistatic PRF} is a necessary condition for the surface of the maximum bistatic range to be smaller than the bistatic surface of the trailing edge of the transmitted pulse $\forall(\vartheta,\varpi)\in[0,2\pi)\times[0,2\pi)$ in~\eqref{eq:Cassini contour}. 

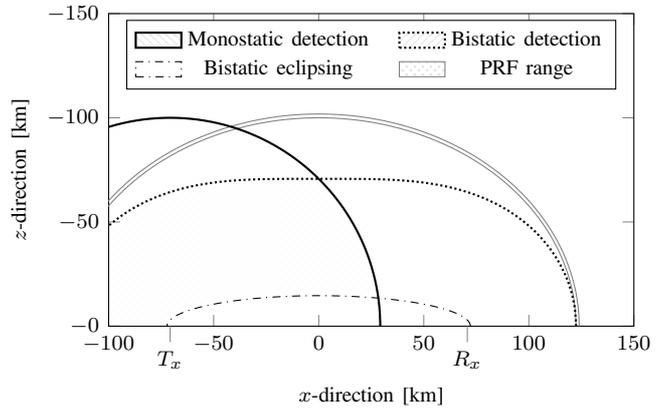
\begin{figure}[!t]%
\definecolor{matlabblue}{rgb}{0,0.4470,0.7410}%
\definecolor{matlabred}{rgb}{0.8500, 0.3250, 0.0980}%
\definecolor{matlabyellow}{rgb}{0.9290, 0.6940, 0.1250}%
\definecolor{matlabmagenta}{rgb}{0.4940, 0.1840, 0.5560}%
\begin{tikzpicture}[%
font=\footnotesize,
dot/.style    = {anchor=base,fill,circle,inner sep=2pt}]		
\begin{axis}[%
width=0.78\columnwidth,
unit vector ratio*=1 1 1,
scale only axis,
xmin=-100,
xmax=150,
xlabel={$x$-direction [km]},
ymin=0,
ymax=150,
ylabel={$z$-direction [km]},
yticklabel=$-\pgfmathprintnumber{\tick}$,
samples=50,
extra x ticks={-70.71, 70.71},
extra x tick labels={$T_x$, $R_x$},
extra x tick style={%
	grid=none, 
	tick align=outside,
  tick pos=left,
	major tick length=0.50\baselineskip},
legend style={%
	at={(0.5,0.98)},
	anchor=north,
	legend columns=2,
	/tikz/every even column/.append style={column sep=0.3cm}},
]

		%

	\addplot[gray,domain=0:180,smooth,variable=\x, name path=PRF1,forget plot] ({123.97*cos(\x)}, {101.83 * sin(\x)});
	\addplot[gray,domain=0:180,smooth,variable=\x, name path=PRF2,forget plot] ({122.47*cos(\x)}, {100.00 * sin(\x)});
	
	\tikzfillbetween[of=PRF1 and PRF2]{opacity=0.1, pattern=crosshatch dots};

	
	\addplot[thick,domain=0:180,smooth,variable=\x, name path=monostatic,forget plot] ({100.00*cos(\x)-70.71}, {100.00 * sin(\x)});
	\addplot[dashdotted,domain=0:180,smooth,variable=\x, name path=blind_range,forget plot] ({72.21*cos(\x)}, {14.64 * sin(\x)});
	
	
	
  \addplot[thick, densely dotted,domain=0:180,smooth,variable=\x, name path=bistatic,forget plot] ({cos(\x) * sqrt(70.71^2 * ( cos(2*\x) + sqrt( (100/70.71)^4 - sin(2*\x)^2  )))}, {sin(\x) * sqrt(70.71^2 * ( cos(2*\x) + sqrt( (100/70.71)^4 - sin(2*\x)^2  )))});
	
	\tikzfillbetween[of=blind_range and bistatic]{opacity=0.1,pattern=north east lines};
	\tikzfillbetween[of=blind_range and monostatic, soft clip={domain=-200:29.29}]{opacity=0.1,pattern=north west lines};

	\addlegendimage{area legend, thick, pattern=north west lines, pattern color=black!10} 
	\addlegendentry{Monostatic detection}
	\addlegendimage{area legend, densely dotted, thick, fill=black!10, pattern=north east lines, pattern color=black!10}
	\addlegendentry{Bistatic detection}
	\addlegendimage{area legend, dashdotted,draw=black, fill=white}
	\addlegendentry{Bistatic eclipsing}
	\addlegendimage{area legend,draw=gray, fill=black!10, pattern=crosshatch  dots, pattern color=black!10}
	\addlegendentry{PRF range}
\end{axis}
\end{tikzpicture}%
\caption{The monostatic detection range, bistatic detection range of~\eqref{eq:Cassini contour}, and bistatic PRF range of~\eqref{eq:PRF ellipse} for $\tau_p=10\mu s$, $R_{bi}=100\mathrm{km}$, $L=\sqrt{2}\cdot100\mathrm{km}$, and the $\PRF_{\mathrm{bi}}=2.81\mathrm{kHz}$.}%
\label{fig:ranges of bistatic conf}%
\vspace*{-6mm}%
\end{figure}

Note that, due to the bistatic configuration, the $\PRF_{\mathrm{bi}}$ in~\eqref{eq:bistatic PRF} can be significantly higher than the monostatic $\PRF_\mathrm{mo}$ with $R_{\mathrm{max}}=L+\Delta R_{\mathrm{bi}}$. As the transmitter and receiver are not co-located, the next pulse can already be transmitted before the reflection of the previous pulse enters the receiver, as~\eqref{eq:bistatic PRF} dictates. If the receiver and transmitter are co-located, then $\left.\Delta R_{\mathrm{bi}}\right|_{L=0}=\sqrt{R_TR_R}$ becomes the monostatic range and the monostatic $\PRF_\mathrm{mo}$ and bistatic $\PRF_{\mathrm{bi}}$ will coincide. An example of the leading ellipse~\eqref{eq:PRF ellipse max} and trailing ellipse~\eqref{eq:PRF ellipse min} for a given $\PRF_{\mathrm{bi}}$ are depicted in Fig.~\ref{fig:ranges of bistatic conf}. In this case, the upper bound $\PRF_{\mathrm{bi}}=2.81\mathrm{kHz}$ is almost doubled w.r.t. the upper bound $\PRF_{\mathrm{mo}}=1.52\mathrm{kHz}$.

The $\PRF_{\mathrm{bi}}$ in~\eqref{eq:bistatic PRF} is significantly lower than the bistatic $\PRF$ reported in~\cite[Fig. 10]{Jackson1986}. In~\cite{Jackson1986}, the high $\PRF$ is achieved by decreasing the unambiguous range to the (small) area of the intersection of the transmit beam and a single, stationary receiver beam. This corresponds to a tracking scenario where the position of the target is already known. Instead, we consider a surveillance scenario without ambiguities where the position of the target is unknown a priori. Furthermore, these high $\PRF$s are only possible if the transmitter is isolated from the receiver.

\section{Pulse Chasing} \label{sec:pulse chasing}

Pulse chasing by the receiver is discussed in this section. First, the required beam switching rate is analysed in Section~\ref{subsec:beam switching rate}. In Section~\ref{subsec:pulse correction}, the pulse search area correction is explained followed by the calculation of the number of required receive beams in Section~\ref{subsec:receiver beamformer}.

\subsection{Beam Switching Rate} \label{subsec:beam switching rate}
\begin{figure}[!t]%
\input{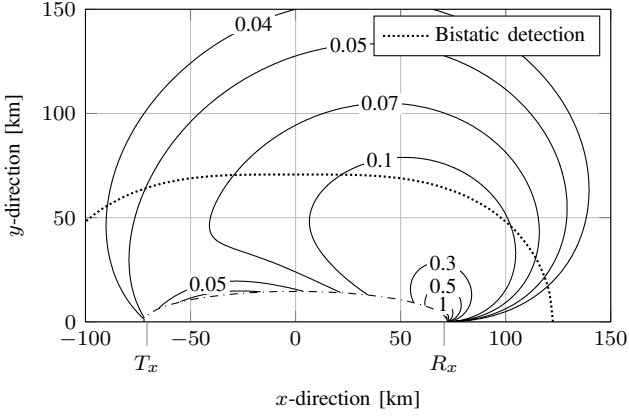}
\caption{Beam switching per $\mu$s for $L=\sqrt{2}\cdot 100\mathrm{km}$ with a beamwidth of $BW_{r,a}=2^\circ$. The maximum rate is $3.395$ beams/$\mu s$.}
\label{fig:switching rate}%
\vspace*{-3mm}%
\end{figure}

The beam switching rate is determined in the azimuth plane of the receiver only. It can straightforwardly been shown that pulse chasing in the azimuth plane leads to the highest beam switching rate~\cite{Purdy2001}. The beam switching rate in the azimuth plane (2D) is~\cite[Eq. (8)]{Purdy2001}:
\begin{equation}
\dot{B}_R = \frac{17.2 L' \sin\varphi_T}{BW_{R,a}\left( \left(R_T'\sin\varphi_T\right)^2 + \left(L'-R_T'\cos\varphi_T\right)^2 \right)},
\label{eq:beam rate}
\end{equation}
where $\dot{B}_R$ is the number of beams to be switched in units of beams per microsecond, $BW_{R,a}\in\mathbb{R}^+$ is the 3dB receiver beamwidth in azimuth, and $R_T',L'$ are in kilometres. The beam switching for the air traffic control radar depicted in Fig.~\ref{fig:ranges of bistatic conf} is given in Fig.~\ref{fig:switching rate}. The figure shows that high switching rates are only seen when the transmit beam passes near the receiver.

\begin{figure}[t!]
    \centering
    \begin{subfigure}[t]{0.40\columnwidth}
        \centering
				\definecolor{mycolor1}{RGB}{0,114,189}%
\definecolor{mycolor2}{RGB}{77,190,238}%
\definecolor{boathull}{RGB}{237,177,32}%
\definecolor{eleAzm}{RGB}{148,83,25}%
\begin{tikzpicture}[x=0.5cm,y=0.5cm,line cap=round,every path/.style={>=latex},scale=0.5]
	\tikzset{dot/.style={circle,fill=#1,inner sep=0,minimum size=4pt}}
	
  \pgfmathsetmacro{\cubebx}{6}
  \pgfmathsetmacro{\cubeby}{5}
  \pgfmathsetmacro{\cubebz}{4}
	
	\pgfmathsetmacro{\cubesx}{3}
  \pgfmathsetmacro{\cubesy}{3}
  \pgfmathsetmacro{\cubesz}{2}
	
	\pgfmathsetmacro{\lengthVirtualLine}{1.5}
	
	\pgfmathsetmacro{\cubesox}{(\cubebx-\cubesx)/2}
  \pgfmathsetmacro{\cubesoy}{(\cubeby-\cubesy)/2}
  \pgfmathsetmacro{\cubesoz}{(\cubebz-\cubesz)/2}

  \draw [draw=black, every edge/.append style={draw=black, densely dashed, opacity=.5}]
    (0,0,0) coordinate (bo) -- ++(-\cubebx,0,0) coordinate (ba) -- ++(0,-\cubeby,0) coordinate (bb) edge coordinate [pos=1] (bg) ++(0,0,-\cubebz)  -- ++(\cubebx,0,0) coordinate (bc) -- cycle
    (bo) -- ++(0,0,-\cubebz) coordinate (bd) -- ++(0,-\cubeby,0) coordinate (be) edge (bg) -- (bc) -- cycle
    (bo) -- (ba) -- ++(0,0,-\cubebz) coordinate (bf) edge (bg) -- (bd) -- cycle;

		
		\draw[->,mycolor1] (bb) -- ($(bb)!0.4!(be)$);
		\draw[-,thick, loosely dotted, mycolor1] ($(bb)!0.4!(be)$) -- (be);
		
		\node[dot=black] at (bo) {};
		\node[dot=black] at (ba) {};
		\node[dot=mycolor2] at (bb) {};
		\node[dot=black] at (bc) {};
		\node[dot=black] at (bd) {};
		\node[dot=black] at (be) {};
		\node[dot=black] at (bf) {};
		\node[dot=black, opacity=.5] at (bg) {};		
		
		%

\end{tikzpicture}%
        \caption{One vertex.}
				\label{fig:relocation vertices 1}
				\vspace*{2mm}
    \end{subfigure}%
    ~ 
    \begin{subfigure}[t]{0.40\columnwidth}
        \centering
				\definecolor{mycolor1}{RGB}{0,114,189}%
\definecolor{mycolor2}{RGB}{77,190,238}%
\definecolor{boathull}{RGB}{237,177,32}%
\definecolor{eleAzm}{RGB}{148,83,25}%
\begin{tikzpicture}[x=0.5cm,y=0.5cm,line cap=round,every path/.style={>=latex},scale=0.5]
	\tikzset{dot/.style={circle,fill=#1,inner sep=0,minimum size=4pt}}
	
  \pgfmathsetmacro{\cubebx}{6}
  \pgfmathsetmacro{\cubeby}{5}
  \pgfmathsetmacro{\cubebz}{4}
	
	\pgfmathsetmacro{\cubesx}{3}
  \pgfmathsetmacro{\cubesy}{3}
  \pgfmathsetmacro{\cubesz}{2}
	
	\pgfmathsetmacro{\lengthVirtualLine}{1.5}
	
	\pgfmathsetmacro{\cubesox}{(\cubebx-\cubesx)/2}
  \pgfmathsetmacro{\cubesoy}{(\cubeby-\cubesy)/2}
  \pgfmathsetmacro{\cubesoz}{(\cubebz-\cubesz)/2}

  \draw [draw=black, every edge/.append style={draw=black, densely dashed, opacity=.5}]
    (0,0,0) coordinate (bo) -- ++(-\cubebx,0,0) coordinate (ba) -- ++(0,-\cubeby,0) coordinate (bb) edge coordinate [pos=1] (bg) ++(0,0,-\cubebz)  -- ++(\cubebx,0,0) coordinate (bc) -- cycle
    (bo) -- ++(0,0,-\cubebz) coordinate (bd) -- ++(0,-\cubeby,0) coordinate (be) edge (bg) -- (bc) -- cycle
    (bo) -- (ba) -- ++(0,0,-\cubebz) coordinate (bf) edge (bg) -- (bd) -- cycle;
		
		\draw[mycolor2] (bc) -- (be);
		
		\draw[->,mycolor1, thick] (bc) -- ($(bc)!0.4!(ba)$);
		\draw[-,thick,loosely dotted, mycolor1] ($(bc)!0.4!(ba)$) -- (ba);
		
		\draw[->,mycolor1, thick] (be) -- ($(be)!0.4!(bf)$);
		\draw[-,thick,loosely dotted, mycolor1] ($(be)!0.4!(bf)$) -- (bf);
		
		\node[dot=black] at (bo) {};
		\node[dot=black] at (ba) {};
		\node[dot=black] at (bb) {};
		\node[dot=mycolor2] at (bc) {};
		\node[dot=black] at (bd) {};
		\node[dot=mycolor2] at (be) {};
		\node[dot=black] at (bf) {};
		\node[dot=black, opacity=.5] at (bg) {};		
		
		%

\end{tikzpicture}%
        \caption{Two vertices.}
				\label{fig:relocation vertices 2}
				\vspace*{2mm}
    \end{subfigure}

		\begin{subfigure}[t]{0.40\columnwidth}
        \centering
				\definecolor{mycolor1}{RGB}{0,114,189}%
\definecolor{mycolor2}{RGB}{77,190,238}%
\definecolor{mycolor3}{RGB}{191,83,25}%
\definecolor{mycolor4}{RGB}{126,47,142}%
\definecolor{boathull}{RGB}{237,177,32}%
\definecolor{eleAzm}{RGB}{148,83,25}%
\begin{tikzpicture}[x=0.5cm,y=0.5cm,line cap=round,every path/.style={>=latex},scale=0.5]
	\tikzset{dot/.style={circle,fill=#1,inner sep=0,minimum size=4pt}}
	
  \pgfmathsetmacro{\cubebx}{6}
  \pgfmathsetmacro{\cubeby}{5}
  \pgfmathsetmacro{\cubebz}{4}
	
	\pgfmathsetmacro{\cubesx}{3}
  \pgfmathsetmacro{\cubesy}{3}
  \pgfmathsetmacro{\cubesz}{2}
	
	\pgfmathsetmacro{\lengthVirtualLine}{1.5}
	
	\pgfmathsetmacro{\cubesox}{(\cubebx-\cubesx)/2}
  \pgfmathsetmacro{\cubesoy}{(\cubeby-\cubesy)/2}
  \pgfmathsetmacro{\cubesoz}{(\cubebz-\cubesz)/2}

  \draw [draw=none, every edge/.append style={draw=none}]
    (0,0,0) coordinate (bo) -- ++(-\cubebx,0,0) coordinate (ba) -- ++(0,-\cubeby,0) coordinate (bb) edge coordinate [pos=1] (bg) ++(0,0,-\cubebz)  -- ++(\cubebx,0,0) coordinate (bc) -- cycle
    (bo) -- ++(0,0,-\cubebz) coordinate (bd) -- ++(0,-\cubeby,0) coordinate (be) edge (bg) -- (bc) -- cycle
    (bo) -- (ba) -- ++(0,0,-\cubebz) coordinate (bf) edge (bg) -- (bd) -- cycle;

		
		\draw[black] (bb) --(ba) -- (bf) -- (bd) -- (bo);
		\draw[black] (bd) --(be);
		\draw[black, densely dashed, opacity=.5] (bb) --(bg) -- (bf);
		
		\draw[mycolor2] (bc) -- (be);
		\draw[mycolor2] (bc) -- (bo);
		
		\draw[->,mycolor1, thick] (bc) -- ($(bc)!0.4!(bb)$);
		\draw[-,thick,loosely dotted, mycolor1] ($(bc)!0.4!(bb)$) -- (bb);
		
		\draw[->,mycolor1, thick] (bo) -- ($(bo)!0.4!(ba)$);
		\draw[-,thick,loosely dotted, mycolor1] ($(bo)!0.4!(ba)$) -- (ba);
		
		\draw[->,mycolor1, thick] (be) -- ($(be)!0.4!(bg)$);
		\draw[-,thick,loosely dotted, mycolor1] ($(be)!0.4!(bg)$) -- (bg);
		
		\node[dot=mycolor2] at (bo) {};
		\node[dot=black] at (ba) {};
		\node[dot=black] at (bb) {};
		\node[dot=mycolor4] at (bc) {};
		\node[dot=black] at (bd) {};
		\node[dot=mycolor2] at (be) {};
		\node[dot=black] at (bf) {};
		\node[dot=black, opacity=.5] at (bg) {};		
		
		%

\end{tikzpicture}%
        \caption{Three vertices.}
				\label{fig:relocation vertices 3}
    \end{subfigure}%
    ~ 
    \begin{subfigure}[t]{0.40\columnwidth}
        \centering
				\definecolor{mycolor1}{RGB}{0,114,189}%
\definecolor{mycolor2}{RGB}{77,190,238}%
\definecolor{mycolor3}{RGB}{191,83,25}%
\definecolor{mycolor4}{RGB}{126,47,142}%
\definecolor{boathull}{RGB}{237,177,32}%
\definecolor{eleAzm}{RGB}{148,83,25}%
\begin{tikzpicture}[x=0.5cm,y=0.5cm,line cap=round,every path/.style={>=latex},scale=0.5]
	\tikzset{dot/.style={circle,fill=#1,inner sep=0,minimum size=4pt}}
	
  \pgfmathsetmacro{\cubebx}{6}
  \pgfmathsetmacro{\cubeby}{5}
  \pgfmathsetmacro{\cubebz}{4}
	
	\pgfmathsetmacro{\cubesx}{3}
  \pgfmathsetmacro{\cubesy}{3}
  \pgfmathsetmacro{\cubesz}{2}
	
	\pgfmathsetmacro{\lengthVirtualLine}{1.5}
	
	\pgfmathsetmacro{\cubesox}{(\cubebx-\cubesx)/2}
  \pgfmathsetmacro{\cubesoy}{(\cubeby-\cubesy)/2}
  \pgfmathsetmacro{\cubesoz}{(\cubebz-\cubesz)/2}

  \draw [draw=none, every edge/.append style={draw=none}]
    (0,0,0) coordinate (bo) -- ++(-\cubebx,0,0) coordinate (ba) -- ++(0,-\cubeby,0) coordinate (bb) edge coordinate [pos=1] (bg) ++(0,0,-\cubebz)  -- ++(\cubebx,0,0) coordinate (bc) -- cycle
    (bo) -- ++(0,0,-\cubebz) coordinate (bd) -- ++(0,-\cubeby,0) coordinate (be) edge (bg) -- (bc) -- cycle
    (bo) -- (ba) -- ++(0,0,-\cubebz) coordinate (bf) edge (bg) -- (bd) -- cycle;

		
		\draw[black] (bb) -- (ba) -- (bf);
		\draw[black, densely dashed, opacity=.5] (bb) --(bg) -- (bf);
		
		\draw[mycolor2] (bo) -- (bc) -- (be) -- (bd) -- (bo);
		
		\draw[->,mycolor1, thick] (bc) -- ($(bc)!0.4!(bb)$);
		\draw[-,thick,loosely dotted, mycolor1] ($(bc)!0.4!(bb)$) -- (bb);
		
		\draw[->,mycolor1, thick] (bo) -- ($(bo)!0.4!(ba)$);
		\draw[-,thick,loosely dotted, mycolor1] ($(bo)!0.4!(ba)$) -- (ba);
		
		\draw[->,mycolor1, thick] (be) -- ($(be)!0.4!(bg)$);
		\draw[-,thick,loosely dotted, mycolor1] ($(be)!0.4!(bg)$) -- (bg);
		
		\draw[->,mycolor1, thick] (bd) -- ($(bd)!0.4!(bf)$);
		\draw[-,thick,loosely dotted, mycolor1] ($(bd)!0.4!(bf)$) -- (bf);
		
		\node[dot=mycolor2] at (bo) {};
		\node[dot=black] at (ba) {};
		\node[dot=black] at (bb) {};
		\node[dot=mycolor2] at (bc) {};
		\node[dot=mycolor2] at (bd) {};
		\node[dot=mycolor2] at (be) {};
		\node[dot=black] at (bf) {};
		\node[dot=black, opacity=.5] at (bg) {};		
		
		%

\end{tikzpicture}%
        \caption{Four vertices.}
				\label{fig:relocation vertices 4}
    \end{subfigure}
    \caption{The blue and magenta vertices represent the to-be-relocated vertices and the arrows indicate the direction of translation.}
		\label{fig:relocation vertices}
		\vspace*{-6mm}
\end{figure}
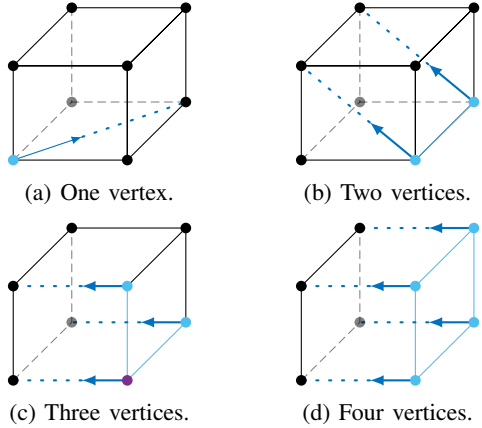

\subsection{Pulse Search Area Correction} \label{subsec:pulse correction}

The vertices $\mathcal{S}_i$ of the box representing the transmitted pulse in~\eqref{eq:3Dbox} are relocated in case that the pulse is partially located inside the eclipsing range~\eqref{eq:eclipsing} to exclude these areas from the calculation of the number of required receiver beams. The centre of the pulse is found at $\mathcal{C}=\mathcal{P}(c(t+\delta t/2), \varphi_T, \beta_T)$ and it should always be outside of the eclipsing range and inside the detection range. Hence, at most four vertices can be inside the eclipsing range. The vertices are translated as follows:
\begin{description}
	\item[1 vertex:] the vertex is translated along the diagonal on the top or bottom plane, i.e., $(x,y)$-plane (Fig.~\ref{fig:relocation vertices 1}).
	\item[2 vertices:] detect the line segment that is connecting the two vertices. The two vertices are translated diagonal on the plane that is perpendicular to the line segment (Fig.~\ref{fig:relocation vertices 2}).
	\item[3 vertices:] detect the vertex that has two connected vertices that are inside the eclipsing range (magenta dot in Fig.~\ref{fig:relocation vertices 3}). The direction of translation is from this vertex to the connected vertex outside the eclipsing range.
	\item [4 vertices:] detect the plane that is spanned by the 4 vertices. The direction of translation of the four vertices is such to compress the box (Fig.~\ref{fig:relocation vertices 4}).
\end{description}
The new location of the vertices will become the intersection point between the line of translation and the eclipsing ellipsoid~\eqref{eq:eclipsing}.

\subsection{Number of receive beams} \label{subsec:receiver beamformer}

\begin{figure}[!t]%
\input{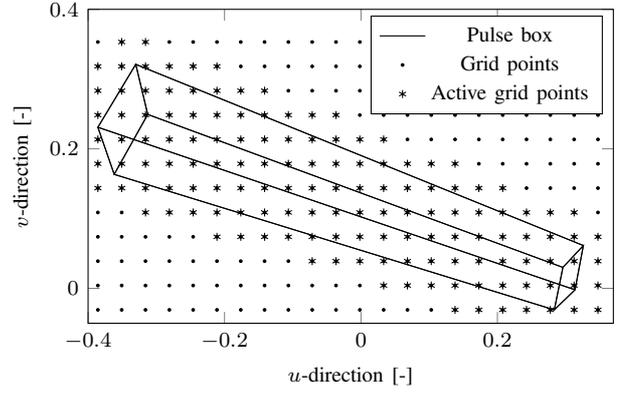}
\caption{Example of the gridding for the receiver multibeam former.}%
\label{fig:gridded pulse}%
\vspace*{-6mm}%
\end{figure}

In this section, the methodology to compute the required number of digital receiver beams using a geometric interpretation is described based on the location of the pulse box discussed in Sections~\ref{subsec:3d coordinate system} and~\ref{subsec:pulse correction}. To determine the number of digital simultaneous beams at the receiver, the location of each vertex coordinate $\mathcal{S}_i$ for $i=\{1,\dots,8\}$ is translated into the $(u,v,w)$-space of the receiver array. The phased array is aligned in $(y,z)$-plane of the receiver. A Cartesian coordinate $\mathcal{P}\in\mathbb{R}^3$ in the receiver body frame corresponds to the following $(u,v,w)$-coordinate
\begin{equation}
\left[\! \begin{array}{ccc} u \!&\! v \!&\! w \end{array} \!\right]^\top = \frac{1}{\left\Vert\mathcal{P}\right\Vert_2}\left[\!\begin{array}{ccc} -\left[ \mathcal{P}\right]_y \!&\! -\left[ \mathcal{P}\right]_z \!&\! \left[ \mathcal{P}\right]_x \end{array} \!\right]^\top,
\label{eq:uvw space}
\end{equation}
where the same Cartesian coordinate in the origin is $\mathcal{P}^\mathcal{O}=\mathcal{P}+\mathcal{R}_x$. By applying the $(u,v,w)$-space, beam broadening for larger aspect angles is automatically included. To obtain the active receiver beams, the $(u,v)$-space is uniformly gridded with $u$ grid space $\sin(BW_{R,a})$ and $v$ grid space $\sin(BW_{R,e})$ where $BW_{R,a},BW_{R,e}\in\mathbb{R}^+$ are the 3dB receiver beamwidths in azimuth and elevation. The grid is started from the minimum in the $(u,v)$ coordinates of the vertices. Then, a grid point is called active if the transmitted beam lies, at least partially, in the rectangle that the point creates with its neighbouring grid points, see Fig.~\ref{fig:gridded pulse}. The number of beams (rectangles) can straightforwardly be found from the active grid points.

\section{Pulse Chasing for a Bistatic Air Traffic Control Radar} \label{sec:pulse chasing results}

In this section, we evaluate the required number of digital beams and the bistatic PRF using a factious bistatic air traffic control radar to detect commercial aircrafts. The simulation setting is described in Section~\ref{subsec:simulation setting}. In Section~\ref{subsec:simulation results}, the simulation results will be discussed.

\subsection{Simulation Setting} \label{subsec:simulation setting}

Our air traffic control radar has a bistatic detection range of $R_\mathrm{bi}=100\mathrm{km}$. To maximize the detection range with respect to the transmitter position, the distance between the transmitter and receiver is chosen $L=\sqrt{2}\cdot100\mathrm{km}$ leading to $\Delta R_\mathrm{bi}=51.76\mathrm{km}$. At this distance, the Cassini oval transitions from a convex shape to a peanut-shape~\cite{Yates1952}. Note that, the receiver is located at a transmitter azimuth of $\phi_T=0^\circ$ and transmitter elevation of $\beta_T=0^\circ$. In the sequel, the pulse length is $\tau_p=10\mu s$, the 3dB beamwidth in azimuth and elevation for both the transmitter and receiver is $BW_{T,a}=BW_{R,a}=BW_{T,e}=BW_{R,e}=2^\circ$, the angle inaccuracy in azimuth and elevation is $\delta\psi_a=\delta\psi_e=1^\circ$, and the timing inaccuracy is $0.5\mu s$.

The search space will be gridded (gridding is discussed next). The centre of the pulse $\mathcal{C}$ is placed at each grid point. If $\mathcal{C}$ is within the eclipsing range, within monostatic range of the transmitter, or outside the bistatic detection range then these grid points will be excluded. In addition, if the transmitted pulse is (partially) above a certain elevation of the receiver, e.g., $70^\circ$, then these points will be excluded (to mimic a real-world phased array). In case that the pulse is partially in front of the array and partially in the back of the array, the phased array is turned $90^\circ$ around the $z$-axis. Hence, the receiver consist out of four phased array panels. Three bistatic scenarios of operation are considered: 
\begin{description}
	\item[Case 1] Pulse chasing in the full search space;
	\item[Case 2] Pulse chasing at the maximum detection range; and
	\item[Case 3] Bistatic tracking.
\end{description}
In case 1, pulse chasing of the full pulse throughout the complete detection volume is obtained. In case 2, the pulse is captured on the maximum detection range only (Cassini oval~\eqref{eq:Cassini contour}) representing a surveillance scenario. For case 3, the size of the box with vertices $\mathcal{S}_i$ in~\eqref{eq:3Dbox} is shrunken to $0.4\mathrm{km}\times0.4\mathrm{km}\times0.4\mathrm{km}$. 

Consider an incoming air plane (e.g., Fig.~\ref{fig:ranges of bistatic conf} at $(150, 10)\mathrm{km}$ with velocity $(-800,0)\mathrm{km/h}$), a more realistic scenario is when the radar detects the plane around the maximum detection range and initializes a target track, i.e., combining the latter two scenarios of operation to decrease the number of simultaneous beams.

For the first case, the transmitted pulse is ``chased'' through the full detection volume. The transmitter azimuth $\phi_T$ and transmitter elevation $\beta_T$ are uniformly gridded between $[-\pi,\pi]$ with $200$ points and $[0,0.5\pi]$ with $50$ points, respectively. In addition, time $t$ is gridded $[0,(L+\Delta R_{bi})/c]$ with grid size of $0.5\mathrm{\mu s}$. For each $(\phi_T,\beta_T)$ pair, the number of receiver beams is computed for all time-steps and the maximum number of receiver beams over all time-steps is stored. 

For the second case, the spherical coordinates of the Cassini surface in~\eqref{eq:Cassini contour} are uniformly gridded on the domain $(\vartheta,\varpi)\in[0,\pi]\times[0,\pi]$ with $200$ points each. The centre of the transmitter pulse $\mathcal{C}$ is placed at the grid point. Thereafter, the associated transmitter azimuth $\varphi_T$, transmitter elevation $\beta_T$, and time instance $t$ can be computed and, hence, the location of the pulse vertices $\mathcal{S}_j$ are known. For each of these spherical coordinates, the number of receiver beams is stored. 

In the third case, the gridding is equivalent to the first case. However, the vertices $\mathcal{S}_i$ of the polygon in~\eqref{eq:3Dbox} are placed such that it defines a cube with size $0.4\mathrm{km}\times0.4\mathrm{km}\times0.4\mathrm{km}$ and centre $\mathcal{C}$ is at the grid point $(\phi_T,\beta_T,t)$.

\subsection{Simulation Results} \label{subsec:simulation results}

Fig.~\ref{fig:PC full area} shows the maximum number of receiver beams for all elevations of the transmitter at a given azimuth direction for pulse chasing (PC) for case 1. The figure includes a case without pulse chasing (wPC), where the receiver covers the full trajectory of the pulse for a given azimuth-elevation pair, and a case where the bistatic eclipsing range~\eqref{subsec:PRF range} is increased to $L+10\mathrm{km}$ (instead of $L+3.00\mathrm{km}$). The figure indicates that following the pulse with the receiver when it travels outward from the transmitter can significantly decreases the number of receiver beams. The reduction in beams is especially evident when the transmitter azimuth is $|\varphi_T|>20^\circ$. In the domain $|\varphi_T|\leq20^\circ$, the number of receiver beams rapidly increases when $|\varphi_T|$ decreases. In this region, the transmitted pulse has significantly widened as it travelled along the major axis of the eclipsing range and it passes the receiver at a very close distance. Similar results are observed by \cite[Sec. 3]{Purdy2001} in 2D. Hence, in the case where the bistatic eclipsing range~\eqref{eq:eclipsing} is increased to $L+10\mathrm{km}$ a decrease in the maximum number of beams is noticed, i.e., from $1600$ to $1444$. In this case, the detection of the pulse is only allowed at a further distance from the receiver, i.e., the receiver beam has significantly widened.

For the second case, Fig.~\ref{fig:PC cassini surface} shows the maximum number of receiver beams for the detection at the maximum bistatic detection range of the radar. In this case, the maximum number of receiver beams is $64$, significantly smaller than the first case.

For the third case, Fig.~\ref{fig:PC target tracking} shows the maximum number of receiver beams to cover a box of $0.4\mathrm{km}\times0.4\mathrm{km}\times0.4\mathrm{km}$ within the bistatic detection range of the radar. In this case, the maximum number of receiver beams is $20$, again significantly smaller than in the first case. This maximum is obtained in case the box is close to the receiver. 

For an adaptive detection paradigm, i.e., combining case 2 and 3, the maximum number of simultaneous receiver beams is $64$, which seems feasible solely from the maximum number of receiver beams and ignoring other hardware and software limitations.

\begin{figure}%
%
\definecolor{matlabblue}{rgb}{0,0.4470,0.7410}%
\definecolor{matlabred}{rgb}{0.8500, 0.3250, 0.0980}%
\definecolor{matlabyellow}{rgb}{0.9290, 0.6940, 0.1250}%
\definecolor{matlabmagenta}{rgb}{0.4940, 0.1840, 0.5560}%
\begin{tikzpicture}[font=\footnotesize]

\begin{polaraxis}[%
width=0.80\columnwidth,
height=0.80\columnwidth,
scale only axis,
rotate=90,
axis line style={gray},
xmin=-90,
xmax=90,
ymax=3.5,
ytick={0,1,2,3},
yticklabels={$10^0$,$10^1$,$10^2$,$10^3$},
xticklabel=$\pgfmathprintnumber{\tick}^\circ$,
    xticklabel style={
        inner xsep=2pt,
        ellipse,
        anchor=\tick-90
    },
    yticklabel style={
        inner xsep=2pt,
        ellipse,
        anchor=\tick+90
    },
ylabel={Number of beams [-]},		
y label style={anchor=north, at={(axis description cs:0.42, -0.5)}},
legend style={anchor=south west, at={(axis cs:-38,3.6)}}
]


\addplot[smooth, matlabblue, line width=1.2, densely dashed] table[x expr=\thisrowno{0},y expr={log10(\thisrowno{1})}, row sep=crcr] {%
%
-90	25\\
-88.20	25\\
-86.40	25\\
-84.60	25\\
-82.80	25\\
-81	27\\
-79.20	29\\
-77.40	29\\
-75.60	29\\
-73.80	31\\
-72.00	31\\
-70.20	29\\
-68.40	32\\
-66.60	33\\
-64.80	33\\
-63	33\\
-61.20	36\\
-59.40	36\\
-57.60	36\\
-55.80	39\\
-54	39\\
-52.20	41\\
-50.40	43\\
-48.60	43\\
-46.80	46\\
-45	50\\
-43.20	50\\
-41.40	55\\
-39.60	58\\
-37.80	63\\
-36	68\\
-34.20	77\\
-32.40	83\\
-30.60	91\\
-28.80	100\\
-27	108\\
-25.20	115\\
-23.40	131\\
-21.60	153\\
-19.80	176\\
-18	202\\
-16.20	227\\
-14.40	253\\
-12.60	284\\
-10.80	327\\
-9	386\\
-7.201	490\\
-5.4	711\\
-3.601	1229\\
-1.8	1653\\
0	966\\
1.8	1653\\
3.601	1229\\
5.4	711\\
7.201	490\\
9	386\\
10.80	327\\
12.60	284\\
14.40	253\\
16.20	227\\
18	202\\
19.80	176\\
21.60	153\\
23.40	131\\
25.20	121\\
27.00	108\\
28.80	98\\
30.60	89\\
32.40	81\\
34.20	74\\
36.00	68\\
37.80	64\\
39.60	58\\
41.40	53\\
43.20	51\\
45.00	50\\
46.80	45\\
48.60	43\\
50.40	43\\
52.20	39\\
54	37\\
55.80	36\\
57.60	35\\
59.40	35\\
61.20	32\\
63	33\\
64.80	31\\
66.60	31\\
68.40	31\\
70.20	32\\
72	28\\
73.80	30\\
75.60	28\\
77.40	27\\
79.20	27\\
81	25\\
82.80	27\\
84.60	25\\
86.40	23\\
88.20	23\\
90	22\\
};
\addlegendentry{wPC};

\addplot[smooth, matlabred, line width=1] table[x expr=\thisrowno{0},y expr={log10(\thisrowno{1})}, row sep=crcr] {%
-43.20	6\\
-41.40	6\\
-39.60	7\\
-37.80	7\\
-36	8\\
-34.20	8\\
-32.40	9\\
-30.60	10\\
-28.80	10\\
-27	12\\
-25.20	23\\
-23.40	24\\
-21.60	32\\
-19.80	34\\
-18	42\\
-16.20	49\\
-14.40	58\\
-12.60	74\\
-10.80	93\\
-9	134\\
-7.201	199\\
-5.40	343\\
-3.601	554\\
-1.80	1060\\
0	1444\\
1.80	1064\\
3.601	554\\
5.40	343\\
7.201	199\\
9	134\\
10.80	93\\
12.60	74\\
14.40	58\\
16.20	49\\
18	42\\
19.80	34\\
21.60	32\\
23.40	24\\
25.20	23\\
27.00	12\\
28.80	11\\
30.60	10\\
32.40	10\\
34.20	9\\
36.00	8\\
37.80	8\\
39.60	6\\
41.40	6\\
43.20	6\\
};
\addlegendentry{PC $10$km};

\addplot[smooth, black, line width=1] table[x expr=\thisrowno{0},y expr={log10(\thisrowno{1})}, row sep=crcr] {%
-43.20	6\\
-41.40	6\\
-39.60	7\\
-37.80	7\\
-36	8\\
-34.20	8\\
-32.40	9\\
-30.60	10\\
-28.80	10\\
-27	12\\
-25.20	23\\
-23.40	24\\
-21.60	32\\
-19.80	34\\
-18	42\\
-16.20	49\\
-14.40	58\\
-12.60	74\\
-10.80	93\\
-9	134\\
-7.201	199\\
-5.4	343\\
-3.601	704\\
-1.8	1163\\
0	1600\\
1.8	1168\\
3.601	704\\
5.4	343\\
7.201	199\\
9	134\\
10.80	93\\
12.60	74\\
14.40	58\\
16.20	49\\
18	42\\
19.80	34\\
21.60	32\\
23.40	24\\
25.20	23\\
27.00	12\\
28.80	11\\
30.60	10\\
32.40	10\\
34.20	9\\
36.00	8\\
37.80	8\\
39.60	6\\
41.40	6\\
43.20	6\\
};
\addlegendentry{PC};

\end{polaraxis}
\end{tikzpicture}%
\caption{The maximum number of receiver beams for a given azimuth direction of the transmitter without pulse chasing (wPC), with pulse chasing (PC), and pulse chasing with a minimal distance of $10\mathrm{km}$ from the receiver (PC $10\mathrm{km}$). The maximum number of beams for wPC is $1653$ , for PC is $1600$, and PC $10\mathrm{km}$ is $1444$.}%
\label{fig:PC full area}%
\end{figure}
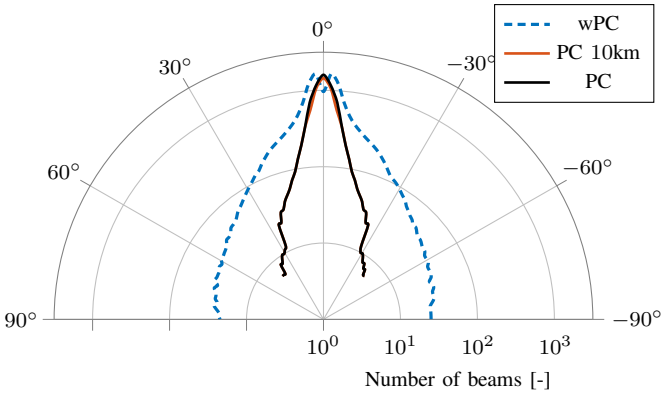

\begin{figure}%
%
\definecolor{matlabblue}{rgb}{0,0.4470,0.7410}%
\definecolor{matlabred}{rgb}{0.8500, 0.3250, 0.0980}%
\definecolor{matlabyellow}{rgb}{0.9290, 0.6940, 0.1250}%
\definecolor{matlabmagenta}{rgb}{0.4940, 0.1840, 0.5560}%
\begin{tikzpicture}[font=\footnotesize]%
\begin{polaraxis}[%
width=0.80\columnwidth,
height=0.80\columnwidth,
scale only axis,
rotate=90,
axis line style={gray},
xmin=-90,
xmax=90,
ymax=70,
xticklabel=$\pgfmathprintnumber{\tick}^\circ$,
    xticklabel style={
        inner xsep=2pt,
        ellipse,
        anchor=\tick-90
    },
    yticklabel style={
        inner xsep=2pt,
        ellipse,
        anchor=\tick+90
    },
ylabel={Number of beams [-]},		
y label style={anchor=north, at={(axis description cs:0.43, -0.5)}},
]


\addplot[smooth, black, line width=1.5, forget plot] table[row sep=crcr] {%
-44.6616541353384	6\\
-43.7593984962406	6\\
-42.8571428571429	6\\
-41.9548872180451	6\\
-41.0526315789474	6\\
-40.1503759398496	6\\
-39.2481203007519	7\\
-38.3458646616541	7\\
-37.4436090225564	7\\
-36.5413533834586	8\\
-35.6390977443609	8\\
-34.7368421052632	8\\
-33.8345864661654	8\\
-32.9323308270677	8\\
-32.0300751879699	8\\
-31.1278195488722	9\\
-30.2255639097744	10\\
-29.3233082706767	10\\
-28.4210526315790	10\\
-27.5187969924812	11\\
-26.6165413533835	18\\
-25.7142857142857	23\\
-24.8120300751880	23\\
-23.9097744360902	24\\
-23.0075187969925	27\\
-22.1052631578947	30\\
-21.2030075187970	32\\
-20.3007518796993	32\\
-19.3984962406015	29\\
-18.4962406015038	27\\
-17.5939849624060	27\\
-16.6917293233083	25\\
-15.7894736842105	29\\
-14.8872180451128	32\\
-13.9849624060150	35\\
-13.0827067669173	39\\
-12.1804511278195	42\\
-11.2781954887218	47\\
-10.3759398496240	48\\
-9.47368421052633	51\\
-8.57142857142858	56\\
-7.66917293233084	56\\
-6.76691729323309	57\\
-5.86466165413535	61\\
-4.96240601503760	64\\
-4.06015037593986	64\\
-3.15789473684211	64\\
-2.25563909774436	64\\
-1.35338345864662	64\\
-0.451127819548873	64\\
0.451127819548873	64\\
1.35338345864662	64\\
2.25563909774436	64\\
3.15789473684211	64\\
4.06015037593986	64\\
4.96240601503760	64\\
5.86466165413535	64\\
6.76691729323309	60\\
7.66917293233084	56\\
8.57142857142858	56\\
9.47368421052633	53\\
10.3759398496240	49\\
11.2781954887218	48\\
12.1804511278195	43\\
13.0827067669173	40\\
13.9849624060150	36\\
14.8872180451128	33\\
15.7894736842105	29\\
16.6917293233083	25\\
17.5939849624060	27\\
18.4962406015038	27\\
19.3984962406015	29\\
20.3007518796993	32\\
21.2030075187970	32\\
22.1052631578947	30\\
23.0075187969925	27\\
23.9097744360902	23\\
24.8120300751880	23\\
25.7142857142857	23\\
26.6165413533835	18\\
27.5187969924812	10\\
28.4210526315790	11\\
29.3233082706767	11\\
30.2255639097744	10\\
31.1278195488722	9\\
32.0300751879699	9\\
32.9323308270677	9\\
33.8345864661654	9\\
34.7368421052632	9\\
35.6390977443609	8\\
36.5413533834586	8\\
37.4436090225564	7\\
38.3458646616541	8\\
39.2481203007519	7\\
40.1503759398496	6\\
41.0526315789474	6\\
41.9548872180451	6\\
42.8571428571429	6\\
43.7593984962406	6\\
44.6616541353384	5\\
};

\end{polaraxis}%
\end{tikzpicture}%
\caption{The maximum number of receiver beams for a given azimuth direction of the transmitter when searching on the maximum detection surface, i.e., the Cassinian surface~\eqref{eq:Cassini contour}. For this case, the maximum number of beams is $64$.}%
\label{fig:PC cassini surface}%
\end{figure}
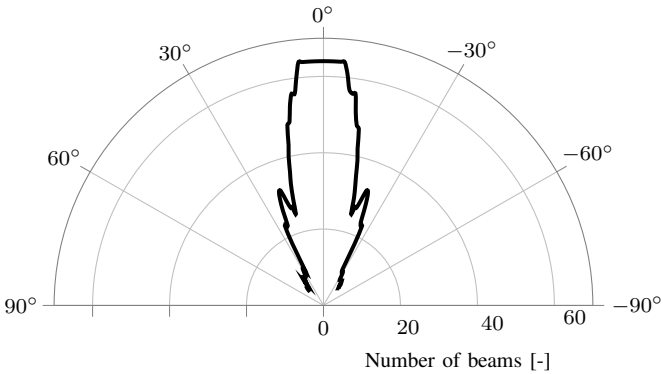

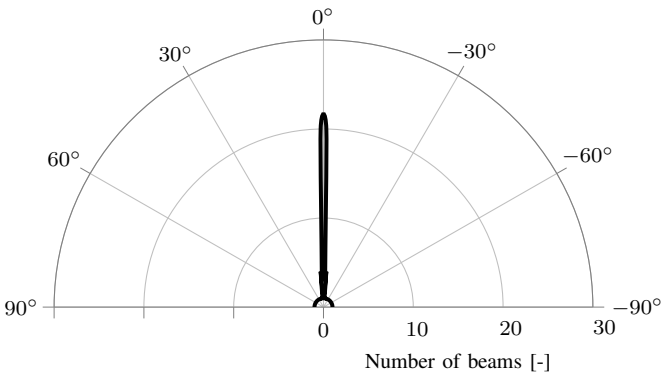
\begin{figure}%
%
\definecolor{matlabblue}{rgb}{0,0.4470,0.7410}%
\definecolor{matlabred}{rgb}{0.8500, 0.3250, 0.0980}%
\begin{tikzpicture}[font=\footnotesize]

\begin{polaraxis}[%
width=0.80\columnwidth,
height=0.80\columnwidth,
scale only axis,
rotate=90,
axis line style={gray},
xmin=-90,
xmax=90,
ymax=30,
xticklabel=$\pgfmathprintnumber{\tick}^\circ$,
    xticklabel style={
        inner xsep=2pt,
        ellipse,
        anchor=\tick-90
    },
    yticklabel style={
        inner xsep=2pt,
        ellipse,
        anchor=\tick+90
    },
ylabel={Number of beams [-]},		
y label style={anchor=north, at={(axis description cs:0.43, -0.5)}},
]


\addplot[smooth, black, line width=1.5, forget plot] table[row sep=crcr] {%
%
-89.5477386934673	1\\
-87.7386934673367	1\\
-85.9296482412060	1\\
-84.1206030150754	1\\
-82.3115577889447	1\\
-80.5025125628141	1\\
-78.6934673366834	1\\
-76.8844221105528	1\\
-75.0753768844221	1\\
-73.2663316582915	1\\
-71.4572864321608	1\\
-69.6482412060302	1\\
-67.8391959798995	1\\
-66.0301507537688	1\\
-64.2211055276382	1\\
-62.4120603015075	1\\
-60.6030150753769	1\\
-58.7939698492462	1\\
-56.9849246231156	1\\
-55.1758793969849	1\\
-53.3668341708543	1\\
-51.5577889447236	1\\
-49.7487437185930	1\\
-47.9396984924623	1\\
-46.1306532663317	1\\
-44.3216080402010	1\\
-42.5125628140704	1\\
-40.7035175879397	1\\
-38.8944723618091	1\\
-37.0854271356784	1\\
-35.2763819095477	1\\
-33.4673366834171	1\\
-31.6582914572864	1\\
-29.8492462311558	1\\
-28.0402010050251	1\\
-26.2311557788945	1\\
-24.4221105527638	1\\
-22.6130653266332	1\\
-20.8040201005025	1\\
-18.9949748743718	1\\
-17.1859296482412	1\\
-15.3768844221106	1\\
-13.5678391959799	1\\
-11.7587939698493	1\\
-9.94974874371858	1\\
-8.14070351758793	1\\
-6.33165829145728	1\\
-4.52261306532663	4\\
-2.71356783919599	4\\
-0.904522613065312	20\\
0.904522613065337	20\\
2.71356783919599	4\\
4.52261306532663	4\\
6.33165829145728	1\\
8.14070351758796	1\\
9.94974874371861	1\\
11.7587939698493	1\\
13.5678391959799	1\\
15.3768844221106	1\\
17.1859296482412	1\\
18.9949748743719	1\\
20.8040201005025	1\\
22.6130653266332	1\\
24.4221105527638	1\\
26.2311557788945	1\\
28.0402010050251	1\\
29.8492462311558	1\\
31.6582914572864	1\\
33.4673366834171	1\\
35.2763819095477	1\\
37.0854271356784	1\\
38.8944723618091	1\\
40.7035175879397	1\\
42.5125628140704	1\\
44.3216080402010	1\\
46.1306532663317	1\\
47.9396984924623	1\\
49.7487437185930	1\\
51.5577889447236	1\\
53.3668341708543	1\\
55.1758793969849	1\\
56.9849246231156	1\\
58.7939698492462	1\\
60.6030150753769	1\\
62.4120603015075	1\\
64.2211055276382	1\\
66.0301507537689	1\\
67.8391959798995	1\\
69.6482412060302	1\\
71.4572864321608	1\\
73.2663316582915	1\\
75.0753768844221	1\\
76.8844221105528	1\\
78.6934673366834	1\\
80.5025125628141	1\\
82.3115577889447	1\\
84.1206030150754	1\\
85.9296482412060	1\\
87.7386934673367	1\\
89.5477386934673	1\\
};

\end{polaraxis}
\end{tikzpicture}%
\caption{The maximum number of receiver beams for a given azimuth direction of the transmitter to cover a cube with an edge size of $0.4\mathrm{km}$. The maximum number of beams is $20$.}%
\label{fig:PC target tracking}%
\end{figure}

\section{Conclusion} \label{sec:conclusion}

In this paper, we have analysed pulse chasing by the receiver in a three dimensional domain considering range, azimuth, and elevation. Namely, a simulation setting is presented to obtain the maximum number of simultaneous receiver beams and the beam switching rate of the digital multidimensional beam formers for various scenarios. In addition, we derive the PRF for the bistatic configuration and it is shown that the bistatic PRF can be increased compared to its monostatic counterpart. Therefore, the unambiguous velocity can be increased without compromising the detection range. The defined notions are applied to analyse an air traffic control radar example. It is shown that an adaptive paradigm can lead to a comprehensible number of simultaneous receiver beams for the bistatic pulse chasing setting.

The current analysis ignores any hardware constrains and hardware inaccuracies, such as limitations on time synchronization, maximum beam switching rate, etc. Including these effects into the analysis will be subject for future research as well as experimental verification of the presented analysis.

\bibliographystyle{IEEEtran}
\def\bibfont{\footnotesize}  
\bibliography{../library}


\appendix
In this appendix, we prove that the surface of the maximum bistatic range will be smaller than the bistatic surface of the trailing edge of the transmitted pulse. More specific, the Cassinian surface~\eqref{eq:Cassini contour} should be within the trailing ellipse~\eqref{eq:PRF ellipse min} for a $\PRF$ satisfying~\eqref{eq:bistatic PRF} and a maximum distance between the transmitter and receiver of $L<2\sqrt{R_TR_R}$. To prove this, it is sufficient to show that the minimum of the radial coordinate of the Cassinian surface~\eqref{eq:Cassini contour} should be smaller than the semi-minor axis of the ellipse~\eqref{eq:PRF ellipse min}, i.e.,  \vspace*{-2mm}
\begin{multline}
\frac{L}{2} \sqrt{\!C\left((k+0.5)\pi\right)}\!-\!\frac{L}{2} = \frac{1}{2}\sqrt{4R_TR_R-L^2} \leq \\ 
\sqrt{ \left( \frac{1}{2} \left(L+c\left(\frac{1}{\PRF}-\tau_p \right)\right) \right)^2 - \left(\frac{L}{2}\right)^2},
\label{eq:Cass small ellip}  \vspace*{-2mm}
\end{multline}
for $k\in\mathbb{Z}$. With simple mathematical manipulations, Eq.~\eqref{eq:Cass small ellip} simplifies to \vspace*{-2mm}
\begin{equation}
\PRF \leq \frac{c}{2\sqrt{R_RR_T}-L+c\tau_p}.
\label{eq:Cass small ellip2} \vspace*{-1mm}
\end{equation}
It can be straightforwardly seen that the denominator of~\eqref{eq:bistatic PRF} is always bigger than~\eqref{eq:Cass small ellip2}, implying that the $\PRF$ upper bound of~\eqref{eq:bistatic PRF} is always smaller than the upper bound of~\eqref{eq:Cass small ellip2}. Hence, obeying~\eqref{eq:bistatic PRF} will automatically satisfy~\eqref{eq:Cass small ellip2} and the maximum bistatic range will be smaller than the bistatic surface of the trailing edge of the transmitted pulse~\eqref{eq:PRF ellipse min}.

\end{document}


\title{Errata: Analysing Multibeam, Cooperative, Ground Based Radar in a Bistatic Configuration%
}

\author{\IEEEauthorblockN{Pepijn B. Cox and Wim L. van Rossum}
\IEEEauthorblockA{Radar Technology, TNO\\
The Hague, The Netherlands \\
Email: \{pepijn.cox, wim.vanrossum\}@tno.nl}}

\maketitle

\begin{abstract}
This errata relates to a correction of~\cite[Fig. 7]{Cox2020}.
\end{abstract}


\IEEEpeerreviewmaketitle

\section{Errata}

In~\cite[Sec. V]{Cox2020}, the authors apply a gridding of the azimuth and elevation space to compute the maximum number of simultaneous receiver beams. The authors recognize that the maximum number of beams for case 3, i.e., the bistatic tracking case, is at an azimuth angle of $\varphi_T=0^\circ$. Unfortunately, \textit{only} for case 3, the grid has been generated by the following MATLAB code%
\begin{lstlisting}
N_azi = 200;
azi_grid = linspace(-pi, pi, N_azi);
\end{lstlisting}
which excludes $\varphi_T=0^\circ$. Missing $\varphi_T=0^\circ$ led to a deviation in the maximum number of simultaneous receiver beams. Hence, the errata is only applicable to~\cite[Fig. 7]{Cox2020} for an azimuth angle between $\varphi_T\in[-0.9045^\circ,-0.9045^\circ]$.\\

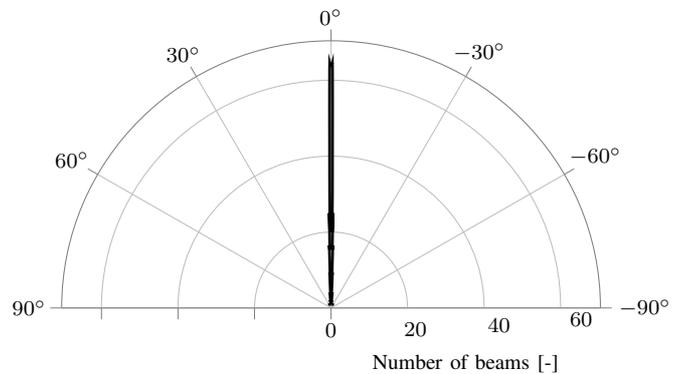
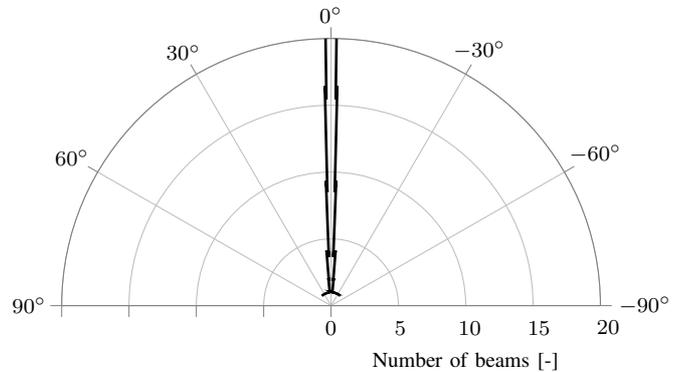
\begin{figure}[!b]%
\centering
	\begin{subfigure}[t]{1\columnwidth}
%
\definecolor{matlabblue}{rgb}{0,0.4470,0.7410}%
\definecolor{matlabred}{rgb}{0.8500, 0.3250, 0.0980}%
\begin{tikzpicture}[font=\footnotesize]

\begin{polaraxis}[%
width=0.80\columnwidth,
height=0.80\columnwidth,
scale only axis,
rotate=90,
axis line style={gray},
xmin=-90,
xmax=90,
xticklabel=$\pgfmathprintnumber{\tick}^\circ$,
    xticklabel style={
        inner xsep=2pt,
        ellipse,
        anchor=\tick-90
    },
    yticklabel style={
        inner xsep=2pt,
        ellipse,
        anchor=\tick+90
    },
ylabel={Number of beams [-]},		
y label style={anchor=north, at={(axis description cs:0.43, -0.5)}},
]


\addplot[smooth, black, line width=1, forget plot] table[row sep=crcr] {
-43.20	1 \\
-41.40	1 \\
-39.60	1 \\
-37.80	1 \\
-36	1 \\
-34.20	1 \\
-32.40	1 \\
-30.60	1 \\
-28.80	1 \\
-27	1 \\
-25.20	1 \\
-23.40	1 \\
-21.60	1 \\
-19.80	1 \\
-18	1 \\
-16.20	1 \\
-14.40	1 \\
-12.60	1 \\
-10.80	1 \\
-9.00000000000005	1\\
-8.81999999999996	1\\
-8.64000000000007	1\\
-8.45999999999998	1\\
-8.27999999999999	1\\
-8.10000000000001	1\\
-7.92000000000002	1\\
-7.74000000000003	1\\
-7.55999999999994	1\\
-7.38000000000005	1\\
-7.19999999999996	1\\
-7.02000000000007	1\\
-6.83999999999998	1\\
-6.65999999999999	1\\
-6.48000000000000	1\\
-6.30000000000002	1\\
-6.12000000000003	1\\
-5.93999999999994	1\\
-5.76000000000005	1\\
-5.57999999999996	1\\
-5.40000000000007	2\\
-5.21999999999998	2\\
-5.03999999999999	2\\
-4.86000000000000	2\\
-4.68000000000001	4\\
-4.50000000000003	4\\
-4.31999999999994	4\\
-4.14000000000005	4\\
-3.95999999999996	4\\
-3.78000000000007	4\\
-3.59999999999998	4\\
-3.41999999999999	4\\
-3.24000000000000	4\\
-3.06000000000001	4\\
-2.88000000000002	4\\
-2.69999999999993	4\\
-2.52000000000005	4\\
-2.33999999999996	9\\
-2.16000000000007	9\\
-1.97999999999998	9\\
-1.79999999999999	9\\
-1.62000000000000	16\\
-1.44000000000001	16\\
-1.26000000000002	16\\
-1.07999999999993	25\\
-0.900000000000046	20\\
-0.719999999999955	30\\
-0.540000000000068	42\\
-0.359999999999978	63\\
-0.179999999999989	64\\
0	64\\
0.180000000000000	64\\
0.360000000000000	63\\
0.540000000000000	42\\
0.720000000000000	30\\
0.900000000000000	20\\
1.08000000000000	25\\
1.26000000000000	16\\
1.44000000000000	16\\
1.62000000000000	16\\
1.80000000000000	9\\
1.98000000000000	9\\
2.16000000000000	9\\
2.34000000000000	9\\
2.52000000000000	4\\
2.70000000000000	4\\
2.88000000000000	4\\
3.06000000000000	4\\
3.24000000000000	4\\
3.42000000000000	4\\
3.60000000000000	4\\
3.78000000000000	4\\
3.96000000000000	4\\
4.14000000000000	4\\
4.32000000000000	4\\
4.50000000000000	4\\
4.68000000000000	4\\
4.86000000000000	2\\
5.04000000000000	2\\
5.22000000000000	2\\
5.40000000000000	2\\
5.58000000000000	1\\
5.76000000000000	1\\
5.94000000000000	1\\
6.12000000000000	1\\
6.30000000000000	1\\
6.48000000000000	1\\
6.66000000000000	1\\
6.84000000000000	1\\
7.02000000000000	1\\
7.20000000000000	1\\
7.38000000000000	1\\
7.56000000000000	1\\
7.74000000000000	1\\
7.92000000000000	1\\
8.10000000000000	1\\
8.28000000000000	1\\
8.46000000000000	1\\
8.64000000000000	1\\
8.82000000000000	1\\
9	1\\
10.80	1 \\
12.60	1 \\
14.40	1 \\
16.20	1 \\
18	1 \\
19.80	1 \\
21.60	1 \\
23.40	1 \\
25.20	1 \\
27.00	1 \\
28.80	1 \\
30.60	1 \\
32.40	1 \\
34.20	1 \\
36.00	1 \\
37.80	1 \\
39.60	1 \\
41.40	1 \\
43.20	1 \\
};

\end{polaraxis}
\end{tikzpicture}%
		\caption{Full figure.}%
		\label{fig:PC target tracking full}%
	\end{subfigure}%
	\\
	\begin{subfigure}[t]{1\columnwidth}
%
\definecolor{matlabblue}{rgb}{0,0.4470,0.7410}%
\definecolor{matlabred}{rgb}{0.8500, 0.3250, 0.0980}%
\begin{tikzpicture}[font=\footnotesize]

\begin{polaraxis}[%
width=0.80\columnwidth,
height=0.80\columnwidth,
scale only axis,
rotate=90,
axis line style={gray},
xmin=-90,
xmax=90,
ymax=20,
xticklabel=$\pgfmathprintnumber{\tick}^\circ$,
    xticklabel style={
        inner xsep=2pt,
        ellipse,
        anchor=\tick-90
    },
    yticklabel style={
        inner xsep=2pt,
        ellipse,
        anchor=\tick+90
    },
ylabel={Number of beams [-]},		
y label style={anchor=north, at={(axis description cs:0.43, -0.5)}},
]


\addplot[smooth, black, line width=1, forget plot] table[row sep=crcr] {
-43.20	1 \\
-41.40	1 \\
-39.60	1 \\
-37.80	1 \\
-36	1 \\
-34.20	1 \\
-32.40	1 \\
-30.60	1 \\
-28.80	1 \\
-27	1 \\
-25.20	1 \\
-23.40	1 \\
-21.60	1 \\
-19.80	1 \\
-18	1 \\
-16.20	1 \\
-14.40	1 \\
-12.60	1 \\
-10.80	1 \\
-9.00000000000005	1\\
-8.81999999999996	1\\
-8.64000000000007	1\\
-8.45999999999998	1\\
-8.27999999999999	1\\
-8.10000000000001	1\\
-7.92000000000002	1\\
-7.74000000000003	1\\
-7.55999999999994	1\\
-7.38000000000005	1\\
-7.19999999999996	1\\
-7.02000000000007	1\\
-6.83999999999998	1\\
-6.65999999999999	1\\
-6.48000000000000	1\\
-6.30000000000002	1\\
-6.12000000000003	1\\
-5.93999999999994	1\\
-5.76000000000005	1\\
-5.57999999999996	1\\
-5.40000000000007	2\\
-5.21999999999998	2\\
-5.03999999999999	2\\
-4.86000000000000	2\\
-4.68000000000001	4\\
-4.50000000000003	4\\
-4.31999999999994	4\\
-4.14000000000005	4\\
-3.95999999999996	4\\
-3.78000000000007	4\\
-3.59999999999998	4\\
-3.41999999999999	4\\
-3.24000000000000	4\\
-3.06000000000001	4\\
-2.88000000000002	4\\
-2.69999999999993	4\\
-2.52000000000005	4\\
-2.33999999999996	9\\
-2.16000000000007	9\\
-1.97999999999998	9\\
-1.79999999999999	9\\
-1.62000000000000	16\\
-1.44000000000001	16\\
-1.26000000000002	16\\
-1.07999999999993	25\\
-0.900000000000046	20\\
-0.719999999999955	30\\
-0.540000000000068	42\\
-0.359999999999978	63\\
-0.179999999999989	64\\
0	64\\
0.180000000000000	64\\
0.360000000000000	63\\
0.540000000000000	42\\
0.720000000000000	30\\
0.900000000000000	20\\
1.08000000000000	25\\
1.26000000000000	16\\
1.44000000000000	16\\
1.62000000000000	16\\
1.80000000000000	9\\
1.98000000000000	9\\
2.16000000000000	9\\
2.34000000000000	9\\
2.52000000000000	4\\
2.70000000000000	4\\
2.88000000000000	4\\
3.06000000000000	4\\
3.24000000000000	4\\
3.42000000000000	4\\
3.60000000000000	4\\
3.78000000000000	4\\
3.96000000000000	4\\
4.14000000000000	4\\
4.32000000000000	4\\
4.50000000000000	4\\
4.68000000000000	4\\
4.86000000000000	2\\
5.04000000000000	2\\
5.22000000000000	2\\
5.40000000000000	2\\
5.58000000000000	1\\
5.76000000000000	1\\
5.94000000000000	1\\
6.12000000000000	1\\
6.30000000000000	1\\
6.48000000000000	1\\
6.66000000000000	1\\
6.84000000000000	1\\
7.02000000000000	1\\
7.20000000000000	1\\
7.38000000000000	1\\
7.56000000000000	1\\
7.74000000000000	1\\
7.92000000000000	1\\
8.10000000000000	1\\
8.28000000000000	1\\
8.46000000000000	1\\
8.64000000000000	1\\
8.82000000000000	1\\
9	1\\
10.80	1 \\
12.60	1 \\
14.40	1 \\
16.20	1 \\
18	1 \\
19.80	1 \\
21.60	1 \\
23.40	1 \\
25.20	1 \\
27.00	1 \\
28.80	1 \\
30.60	1 \\
32.40	1 \\
34.20	1 \\
36.00	1 \\
37.80	1 \\
39.60	1 \\
41.40	1 \\
43.20	1 \\
};

\end{polaraxis}
\end{tikzpicture}%
		\caption{Zoomed.}%
		\label{fig:PC target tracking zoom}%
	\end{subfigure}%
	\caption{The maximum number of receiver beams for a given azimuth direction of the transmitter to cover a cube with an edge size of $0.4\mathrm{km}$. The maximum number of beams is $64$.}
	\label{fig:PC target tracking}
\end{figure}

Figure~\ref{fig:PC target tracking} is created based on the same setting as for \cite[Fig. 7]{Cox2020} where \mcode{N_azi = 201} to include $\varphi_T=0^\circ$. Similar to~\cite[Fig. 7]{Cox2020}, in Figure~\ref{fig:PC target tracking}, a large spike can be seen around $\varphi_T=0^\circ$ with a maximum of $64$ simultaneous receiver beams, instead of $20$ previously reported in~\cite{Cox2020}. Even on these new insights, the tracking setting still has an acceptable number of maximum simultaneous receiver beams. Hence, the adaptive paradigm, i.e., combining surveillance on the maximum detection range combined with bistatic tracking, can lead to a comprehensible number of simultaneous receiver beams for the bistatic pulse chasing setting.

\bibliographystyle{IEEEtran}
\def\bibfont{\footnotesize}  
\bibliography{../library}
